\documentclass[prb]{revtex4}
\usepackage[dvips]{graphicx}
\usepackage{epsfig}
\usepackage{amsmath,amsfonts,amssymb,bm}

\begin{document}
\title{Low temperature dipolar echo in amorphous dielectrics: Significance of relaxation and decoherence free two level systems.}
\author{Alexander L. Burin, and John M. Leveritt III,}
\affiliation{Department of Chemistry, Tulane University, New
Orleans, LA 70118, USA}
\author{Gudrun Fickenscher, Andreas Fleischmann, Christian Sch\"otz, Mesoomeh Bazrafshan, Paul Fa\ss l, Manfred v. Schickfus and Christian Enss}
\affiliation{Kirchhoff Institut f\"ur Physik,   
Department of Physics and Astronomy, Heidelberg University, Im Neuenheimer Feld 227
D-69120 Heidelberg, Germany}
\date{\today}
\begin{abstract}
The nature  of dielectric echoes  in amorphous solids at low temperatures is investigated.  It is shown that at long delay times the echo amplitude is determined by a small subset of two level systems (TLS) having negligible relaxation and decoherence because of their weak coupling to phonons. The echo decay can then be described approximately by power law time dependencies with different powers at times longer and shorter than the typical TLS relaxation time. The theory is applied to recent measurements of two and three pulse dipolar echo in borosilicate glass BK7 and provides a perfect data fit in the broad time and temperature ranges under the assumption that there exist two TLS relaxation mechanisms due to TLS-phonons and TLS-TLS interaction. This interpretation is consistent with the previous experimental and theoretical investigations.  Further experiments verifying the theory predictions are suggested.  
\end{abstract}

\maketitle

\begin{figure}[h!]
\centering
\includegraphics[width=8cm]{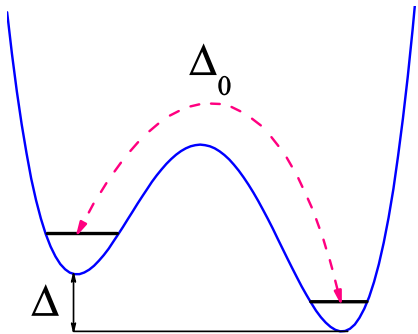}
\caption{Model of TLS. $\Delta$ is the energy difference between left and right well states when isolated, which are coupled with the tunneling amplitude $\Delta_0$.}
\label{Fig1TLS}
\end{figure}

Two level systems (TLS) in low temperature amorphous solids have recently attracted growing attention due to their performance limiting effects in superconducting qubits for quantum computing \cite{Martinec,Ustinov,KO1} and kinetic inductance photon detectors for astronomy \cite{Gao}. TLS are represented by atoms or groups of atoms tunneling between two close energy minima (see Fig. \ref{Fig1TLS}, Refs. \cite{AHVP,Hunklinger}). Understanding of the TLS effect on thermodynamics and kinetics of materials requires knowledge of their dynamical properties including relaxation and decoherence rates. These rates can be determined using the spin-echo technique\cite{nuclear} generalized for amorphous solids.\cite{echo1} However, the time-dependence of echo signals is rather complex and cannot be expressed using the simple exponential decay directly determining TLS relaxation or decoherence rates. Instead, since all TLS interact differently with the external pulse the decay of echo signal as a function of delay time is much more complicated.\cite{echo1,BaierSchickfus,thesis1} To our knowledge there is no general theory capable to describe the complicated echo signal and use it to extract the information about TLS relaxation and decoherence. Below we propose such a theory averaging individual TLS responses over their parameters with the special attention to their coupling to phonons. In our scenario the echo signal is determined by the small subset of relaxation and decoherence free TLS formed due to fluctuations of environment in a certain extent similarly to Ref. \cite{Ora}, where this idea has been proposed and explored for quantum bits (qubits).   A common sense based assumption about TLS coupling statistics results in an almost perfect fit of experimental data for Borosilicate glass BK7 with the fit parameters consistent with the previous experimental data obtained.\cite{Gudrun} In these experiments the echo amplitude has been observed for unprecedentedly long times following its five order of magnitude decay (see Figs. \ref{fig:Fig2TLS} and \ref{fig:Fig3TLS}) which creates necessary grounds for the present theoretical development.

TLS can be characterized by a potential well asymmetry $\Delta$ and a tunneling amplitude $\Delta_{0}$, distributed according to the universal law, $P(\Delta, \Delta_{0})\approx P_{0}/\Delta_{0}$, which reflects the exponential sensitivity of a tunneling amplitude to a potential barrier.  TLS is also coupled  to continuum environment including phonons and other TLS. For each TLS, $i$, its coupling to environment is determined by its elastic tensor  $\widehat{\gamma}_{i}$ describing its interaction with the strain field (phonons), $\widehat{\epsilon}$, given by $-\gamma_{i}^{ab}\epsilon_{ab}S_{i}^{z}$ ($a, b=x, y, z$).
It is convenient to represent the TLS elastic tensor by its invariant  $\mid\widehat{\gamma}\mid\approx\sqrt{{\rm Tr}{\widehat{\gamma}^{2}}}$, having average root mean squared value $\gamma_{0}$ because the limit $|\widehat{\gamma}|\rightarrow 0$, where TLS relaxation and decoherence rates also becomes very small compared to their average values, is relevant for the long time echo amplitude as pointed out below.

For the further consideration one can break the TLS elastic tensor into two parts including the transverse, $\widehat{\gamma'}_{t}=\widehat{\gamma}-\frac{1}{3}{\rm Tr}(\widehat{\gamma})\widehat{I}$ and longitudinal, $\widehat{\gamma'}_{l}=\widehat{I}{\rm Tr}(\widehat{\gamma})$, parts ($\widehat{I}$ is the $3\times 3$ unit matrix). The longitudinal part is coupled only to longitudinal phonons having larger sound velocity, $c_{l}$, than the transverse ones, $c_{t}$. Since the relative weight of the longitudinal contribution to TLS interaction and relaxation decreases with the parameter $c_{t}/c_{l}$ (for instance for the phonon stimulated relaxation it scales as\cite{Jackle72}  $(c_{t}/c_{l})^{5}\ll 1$) one can approximately ignore the longitudinal part of the  elastic tensor  and restrict the consideration to its transverse part (see discussion after Eq. (\ref{eq:EchoesTotalIntegr})).  

TLS interaction with phonons and with each other results in their relaxation to equilibrium which is the subject of interest for echo experiments. According to the earlier theoretical  \cite{Jackle76,abrel,abreview} and experimental \cite{EnssReview,Esquinazi,OsheroffRelaxation,Classen,Ladieu,Fefferman} studies the TLS relaxation rate in dielectric glasses can be represented as
\begin{eqnarray}
k_{i}(E, \Delta_{0}, T)\approx \frac{\Delta_{0}^{2}}{E^{2}}\frac{|\widehat{\gamma}_{i}|^2}{\gamma_{0}^2}k_{s},
\nonumber\\
k_{s}(E, T)=A\left(\frac{E}{k_{B}}\right)^{3}\coth\left(\frac{E}{2k_{B}T}\right) + BT, 
\nonumber\\ 
A \sim \frac{\gamma_{0}^{2}k_{B}^3}{\rho c^{5}\hbar^4}, ~ B \sim 10\frac{(P_{0}U_{0})^{3}}{\hbar}, ~ U_{0} \sim \frac{\gamma_{0}^2}{\rho c^{2}},
\label{eq:DiagRelRate}
\end{eqnarray}
where $E=\sqrt{\Delta^{2}+\Delta_{0}^2}$ is the TLS excitation energy,  $\rho$ is the material density, $c$ is the characteristic speed of sound, and $U_{0}$ is the average absolute value of TLS-TLS elastic interaction constant for $1/R^{3}$ interaction, while their dipole-dipole interaction is usually smaller and neglected here.\cite{Hunklinger,abreview,BlackHalperin} It is convenient to characterize the TLS-TLS interaction constant by its average absolute value, since it enters in that form to the spectral diffusion\cite{BlackHalperin} and relaxation.\cite{abreview} The rate $k_{s}$ describes the relaxation of a ``typical" symmetric ($|\widehat{\gamma}_{i}|=\gamma_{0}$, $\Delta=0$)  TLS.  
The first contribution to the TLS relaxation rate $\left(A\left(\frac{E}{k_{B}}\right)^{3}\coth\left(\frac{E}{2k_{B}T}\right)\right)$ is due to TLS-phonon interaction, while the second one ($BT$) is attributed to the TLS-TLS interaction. Both experiment and theory are qualitatively consistent with Eq. (\ref{eq:DiagRelRate}), but the observed numerical prefactor $B$ is larger by around two orders of magnitude than the related theoretical prediction.\cite{abreview}

The  TLS $i$ interaction with other TLS's is determined by its interaction constant $U_{0i}$ which determines the spectral diffusion and decoherence for the given TLS $i$.\cite{BlackHalperin} In the case of interest where TLS coupling to phonons is small $\mid \widehat{\gamma}_{i} \mid\rightarrow 0$ this interaction scales linearly with $\mid \widehat{\gamma}_{i} \mid$ (therefore the TLS-TLS interaction stimulated relaxation rate scales as $\mid \widehat{\gamma}_{i} \mid^2$ in Eq. (\ref{eq:DiagRelRate}), cf. \cite{abreview}) so we can set $U_{0i}=U_{0}\frac{|\widehat{\gamma}_{i}|}{\gamma_{0}}$.


The TLS contribution to the echo amplitude decreases exponentially with the characteristic times of experiment, $t_{12}$, $t_{13}$ (see Eq. (\ref{eq:echoesTheoryExpAsympt}) below). However, a subset of TLS having very small coupling to phonons, $|\widehat{\gamma}|\ll \gamma_{0}$, can yet strongly contribute to the echo at arbitrarily long times. To describe the response of such subset we introduce the distribution function $F(x)$ for the relative TLS-phonon coupling strength  $x=|\widehat{\gamma}|/\gamma_{0}$. In our model the interaction tensor $\widehat{\gamma}$ is a symmetric $3\times 3$ traceless matrix having $5$ independent components. To have $|\widehat{\gamma}|$ approaching $0$ one needs all these five components to approach zero. This suggests $F(x)dx\propto dx^5$ for $x\ll 1$. At large $x\geq 1$ we expect that its statistics approximately obeys  the Gaussian distribution in agreement with the law of large numbers for the sum of independent contributions. Combining two limiting regimes we propose the distribution function for the relative coupling constant $x$ in the form 
\begin{equation}
F(x)=\frac{25\sqrt{10}}{3\sqrt{\pi}}x^4 e^{-\frac{5x^2}{2}}.
\label{eq:gammadistr}
\end{equation}
This function is chosen to be normalized by unity and to have the average squared value of $x$ to be also equal to unity. 

\begin{figure}[ht]
\centering
\begin{center}$
\begin{array}{cc}
\includegraphics[width=7cm]{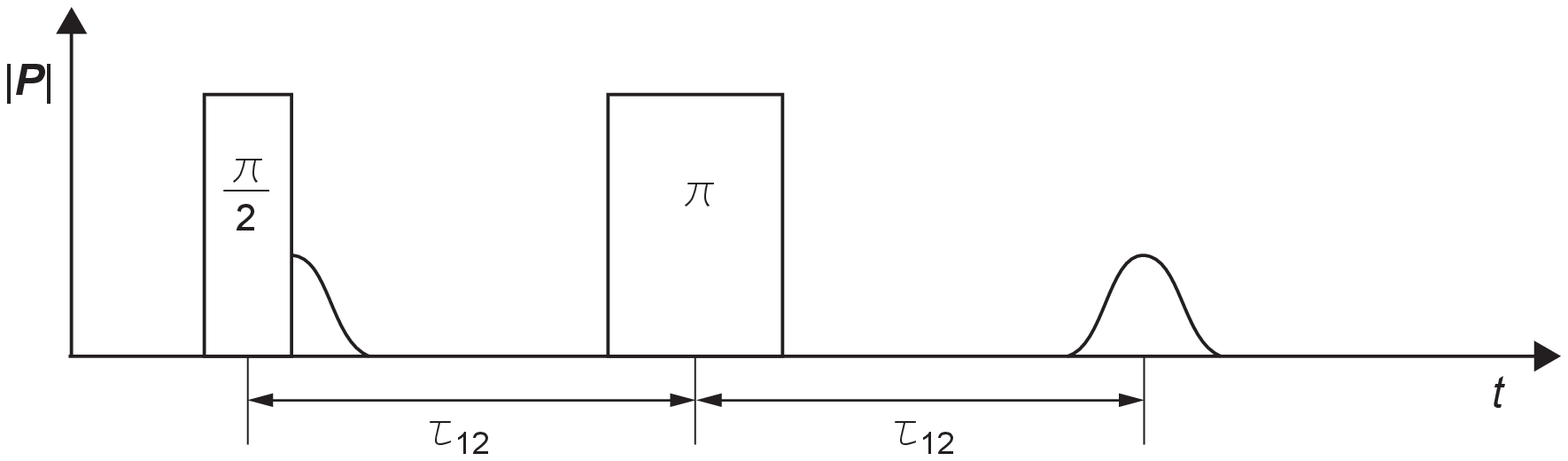} & 
\includegraphics[width=7cm]{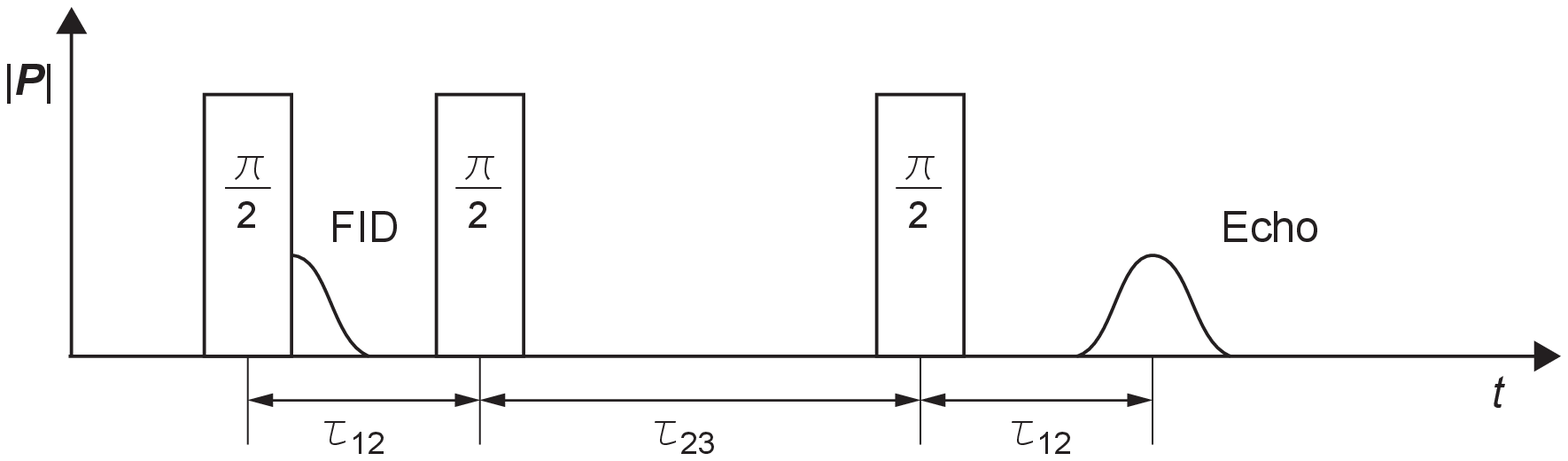} 
\end{array}$
\end{center}
\caption{Two ({\bf left}) and three ({\bf right}) pulse echo sequences}
\label{fig:EchoDefs}
\end{figure}

\begin{figure}[h!]
\centering
\includegraphics[width=8cm]{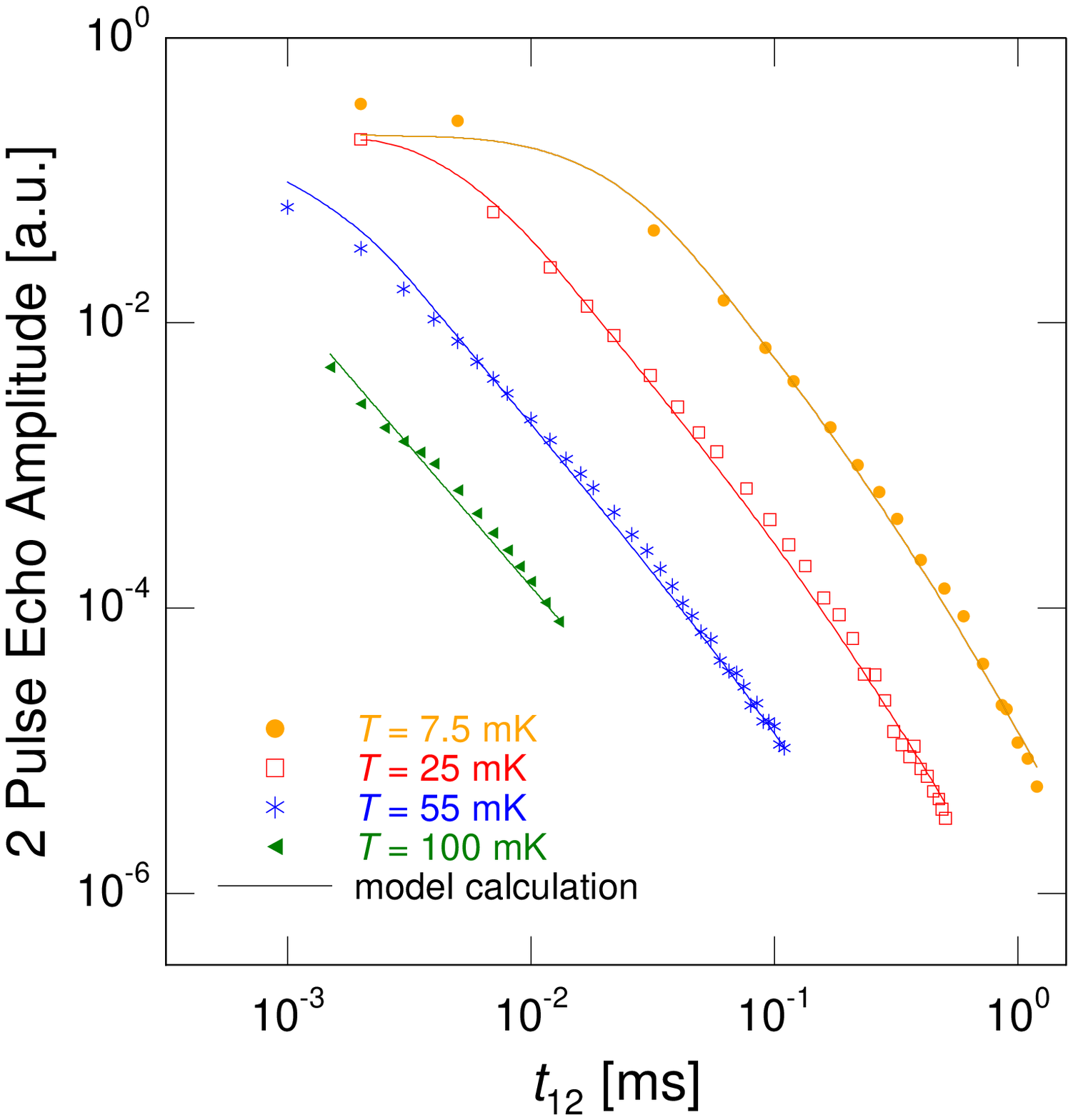}
\caption{ Experiment (symbols) vs. theory (lines) for the time dependence of the two pulse echo amplitude in BK7. Straight lines (colored online) from the top to the bottom corresponds to experimental temperatures in ascending order, $7.5$mK, $25$mK, $55$mK and $100$mK, respectively.}
\label{fig:Fig2TLS}
\end{figure}

\begin{figure}[h!]
\centering
\includegraphics[width=8cm]{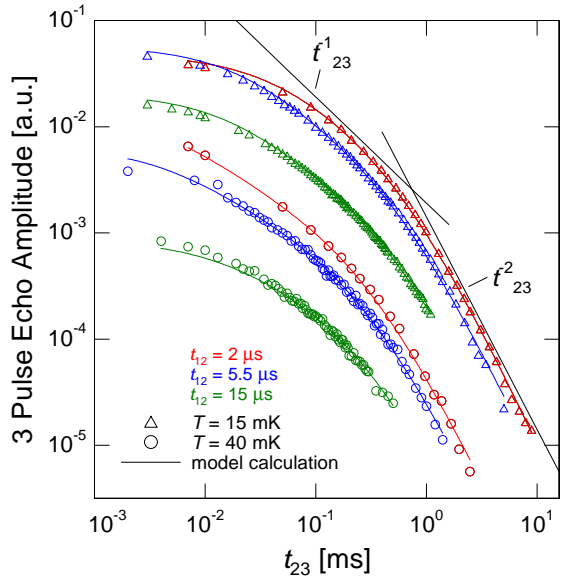}
\caption{ Experiment (symbols) vs. theory (lines) for the time dependence of the three pulse echo signal in BK7. Straight lines (colored on line) from the top to the bottom corresponds to experimental temperatures and times, $t_{12}$, in ascending order, $15$mK ($2\mu$s, $5.5\mu$s, $15\mu$s), $40$mK ($2\mu$s, $5.5\mu$s, $15\mu$s), respectively.}
\label{fig:Fig3TLS}
\end{figure}

Echoes occur when a series of pulses at the same frequency, $\omega_{0}$, are applied to the glassy material. Here we consider dielectric echoes formed by microwave pulses characterized by electric field amplitude, $F$, interacting with TLS dipole moments, $p\sim 1$D. Two pulse echoes are composed by two pulses of duration $\tau$ (``$\pi/2$-pulse") and $2\tau$ (``$\pi$-pulse"), separated by the time $t_{12}$, while the three pulse echoes are composed by three $\pi/2$-pulses, separated by times $t_{12}$ and $t_{23}$ (see Ref. \cite{echo1,BlackHalperin} and Fig. \ref{fig:EchoDefs}).

The echo signal is observed at the frequency $\omega_{0}$ of the pump pulses after the time $t_{12}$ from the last pulse. This signal is caused by the reversed precession of the resonant TLS polarization ($E_{0}=\hbar\omega_{0}$) turned by the $\pi$ angle due to a $\pi$ pulse (two pulse echo) or two $\pi/2$- pulses (three pulse echo, see Refs. \cite{echo1,nuclear,BaierSchickfus}). Actual ``turn angles" of resonant TLS  wavefunction during the pulse, $\varphi=\tau \mathbf{pF}\Delta_{0}/(2E\hbar)$ differ for different TLS and they are not equal to $\pi/2$ or $\pi$. We assume that all turn angles are smaller than $1$ so that the perturbation theory with respect to the external field is applicable, though this is not significant for the echo signal at intermediate and long times. The evolution of TLS wavefunction $(c_1, c_2)$ initially occupying its ground state $c_1=1, c_2=0$ (consideration of excited state simply adds the factor of the equilibrium population difference $\tanh\left(\frac{E_{0}}{2k_{B}T}\right)$) during the two pulse echo sequence can be approximately described as \cite{echo1,BlackHalperin,BaierSchickfus} (cf. Eq. (\ref{eq:DiagRelRate}))
\begin{eqnarray}
\left( \begin{array}{c} 1 \\ 0 \end{array}\right) \underset{\frac{\pi}{2}}{\rightarrow} \left( \begin{array}{c} 1 \\ i\varphi \end{array}\right) \underset{t_{12}}{\rightarrow} \left( \begin{array}{c} e^{i\frac{\Phi_{1}}{2}}  \\ i\varphi e^{-i\frac{\Phi_{1}}{2}-\frac{kt_{12}}{2}}\end{array}\right) \underset{\pi*}{\rightarrow} \left( \begin{array}{c} 2\varphi^2 e^{-i\frac{\Phi_{1}}{2}-\frac{kt_{12}}{2}}\\ i\varphi e^{i\frac{\Phi_{1}}{2}} \end{array}\right) \underset{t_{12}'}{\rightarrow} e^{-\frac{kt_{12}}{2}}\left( \begin{array}{c} 2\varphi^2 e^{-i\frac{\Phi_{1}-\Phi_{2}}{2}}\\ i\varphi e^{i\frac{\Phi_{1}-\Phi_{2}}{2}} \end{array}\right),
\nonumber\\
k=(1-q^2)x^{2}k_{s}(E_{0},T) 
~ x=\frac{|\widehat{\gamma}|}{\gamma_{0}}, ~ q=\frac{|\Delta|}{E}, 
\label{eq:EchoFormation}
\end{eqnarray}
where the phases, $\Phi$, are defined as $\hbar\Phi_{1}(t_{12})=\int_{0}^{t_{12}}d\tau E(\tau)$, $\hbar\Phi_{2}(t_{12}')=\int_{0}^{t_{12}'}d\tau E(\tau+t_{12})$, $E(\tau)$ is the TLS energy, fluctuating with the time due to spontaneous transitions of neighboring TLS. The echo signal can be observed at the time $t_{12}'=t_{12}$  due to the approximate compensation of phases $\Phi_{1}\approx \Phi_{2}$. Its intensity is given by the average TLS dipole moment off-diagonal element, oscillating with the resonant frequency,
\begin{eqnarray}
<p_{2{\rm p}}>\approx p\frac{\Delta_{0}}{2E_{0}}(c_{1}^{*}c_{2}+c_{2}^{*}c_{1})\propto -\sin(\omega t)
(1-q^2)^{2}e^{-k_{res}(1-q^2)x^{2}t_{12}}<e^{i(\Phi_{1}(t_{12})-\Phi_{2}(t_{12}))}>.
\label{eq:Echotwo}
\end{eqnarray}
We assumed that the absolute value of TLS dipole moment is approximately constant, $p$.\cite{Martinec} As it follows from the preliminary analysis this assumption results in the best fit for the echo amplitude dependence  on the pulse time. The discussion of this problem in detail is beyond the scope of this paper. 

Similarly one can obtain the three pulse echo amplitude in the form\cite{BlackHalperin} 
\begin{eqnarray}
<p_{3{\rm p}}>\approx  -\sin(\omega t)(1-q^2)^{2}e^{-k_{res}(1-q^2)x^2(t_{12}+t_{23})}<e^{i(\Phi_{1}(t_{12})-\Phi_{3}(t_{12}, t_{23}))}>,
\nonumber\\
2\hbar\Phi_{3}(t_{12}, t_{23})=\int_{0}^{t_{12}}d\tau E(\tau+t_{23}). 
\label{eq:Echothree}
\end{eqnarray}

Echo amplitudes decay with the delay times, $t_{12}$, $t_{23}$, due to TLS relaxation, Eq. (\ref{eq:DiagRelRate}) and phase decoherence,\cite{BlackHalperin,thesis1,HuHartmann} occurring due to its phase  fluctuations associated with its interaction with neighboring TLS making transitions between their ground and excited states. These fluctuations lead to  the decay of neighbor averaged phase exponents in Eqs. (\ref{eq:Echotwo}), (\ref{eq:Echothree}) as 
\begin{eqnarray}
<e^{i(\Phi_{1}(t_{12})-\Phi_{2}(t_{12}))}>=\exp\left[-xqG_{2{\rm p}}(t_{12})\right], 
\nonumber\\
<e^{i(\Phi_{1}(t_{12})-\Phi_{3}(t_{12}, t_{23}))}>=\exp\left[-xqG_{3{\rm p}}(t_{12}, t_{23})\right].
\label{eq:EchoDecoherenceExponentsDef}
\end{eqnarray}
The factor $q$ appears in the exponents because the TLS ``diagonal'' interaction with neighbors scales as $U\propto \frac{\Delta}{E}$ while the factor $x$ is because the elastic interaction coupling constant is proportional to the absolute value of TLS-phonon interaction tensor (see Ref. \cite{abreview,BlackHalperin}). The linear dependence of exponents in Eq. (\ref{eq:EchoDecoherenceExponentsDef}) of these factors is a consequence of the specific of the TLS $1/R^{3}$ interaction.\cite{BlackHalperin,thesis1,HuHartmann} As it is shown in Sec. \ref{sec:App} the functions $G_{2{\rm p},3{\rm p}}$ can be approximated by the following analytical expressions similar to the widths of the spectral diffusion\cite{BlackHalperin} 
\begin{eqnarray}
G_{2{\rm p}}(t_{12})=\frac{\alpha k_{B}T t_{12}}{\hbar}\frac{\ln\left(1+\eta_{2{\rm p}}k_{T}t_{12}\right)}{\eta_{2{\rm p}}};
\nonumber\\
G_{3{\rm p}}(t_{12}, t_{23})=\frac{\alpha k_{B}Tt_{12}}{\hbar}\frac{\ln\left(1+\eta_{3{\rm p}}k_{T}(t_{12}+t_{23})\right)}{\eta_{3{\rm p}}};
\nonumber\\
k_{T}=AT^{3}+ \xi BT. ~ \alpha=\frac{\pi^{6}}{24} P_{0}U_{0}.
\label{eq:EchoesTheorySingleTLS}
\end{eqnarray}
Here $k_{T}$ is the characteristic relaxation rate of symmetric TLS ($\Delta_{0}=E\sim k_{B}T$, $|\widehat{\gamma}|=\gamma_{0}$), the free parameter $\alpha$ expresses the interaction strength of TLS and the free parameter $\xi$ accounts for the difference between average relaxation rates caused by TLS-TLS and TLS-phonon interactions (see Appendix). At short times $t_{12}$, $t_{23}\ll 1/k_{T}$ we obtain $G\approx \frac{TP_{0}U_{0}k_{T}t_{12}(t_{12}+t_{23})}{\hbar}$ in a full accord with Refs. \cite{BlackHalperin,HuHartmann}. The logarithmic behavior at longer times is due to integrated contribution of neighboring TLS possessing small tunneling amplitudes, $\Delta_{0}\ll k_{B}T$. The parameters $\eta_{2{\rm p}}$, $\eta_{3{\rm p}}$ are used as the free parameters of the theory, describing the transition between two asymptotic time dependence. These parameters are determined below by the direct comparison of theory and experiment. The  relaxation parameters, $A$ and $B$,  Eqs. (\ref{eq:DiagRelRate}), (\ref{eq:EchoFormation}), are determined similarly.

The integrated echo signal is calculated by averaging of Eq. (\ref{eq:EchoesTheorySingleTLS}) over the distribution of TLS tunneling amplitudes and the statistics of their coupling constants, Eq. (\ref{eq:gammadistr}). The final result for the echo signal can be expressed as 
\begin{eqnarray}
I_{2{\rm p},3{\rm p}}\propto\int_{0}^{\infty}x^4 e^{-\frac{5x^2}{2}}dx\int_{0}^{1}dq(1-q^2) e^{-x^2(1-q^{2})k_{res}(t_{12}+t_{23})-xq\frac{\alpha k_{B}Tt_{12}}{\hbar}\frac{\ln\left(1+\eta_{2{\rm p},3{\rm p}}k_{T}(t_{12}+t_{23})\right)}{\eta_{2{\rm p},3{\rm p}}}}. 
\label{eq:EchoesTotalFit}
\end{eqnarray}
For the two pulse echo one should set $t_{23}=0$ in the above expression. 

Consider the  qualitative nature of the echo amplitude time dependence, Eq. (\ref{eq:EchoesTotalFit}), at various observation times. At short times, $t_{12}^2$ or $t_{12}t_{13} < \frac{\hbar}{k_{B}TP_{0}U_{0}k_{T}}$, all TLS contribute with $q$-dependent weight to the signal, decaying by an order of magnitude during times $t_{12}\approx \sqrt{ \frac{\hbar}{k_{B}TP_{0}U_{0}k_{T}}}$ (two pulse echo) or $t_{23}\approx  \frac{\hbar}{k_{B}TP_{0}U_{0}k_{T}t_{12}}$ (three pulse echo) in a qualitative agreement with earlier observations (see Review \cite{PhillipsReview} and references therein). At longer times the echo signal is determined by almost fully symmetric two level systems $\Delta\ll \Delta_{0}\approx E_{0}$, meaning $q\ll 1$ in Eq. (\ref{eq:EchoesTotalFit}). Then one can set $1-q^2\rightarrow 1$ and expand the integral in Eq. (\ref{eq:EchoesTotalFit}) over $q$ to infinity. This yields  
\begin{eqnarray}
I_{2{\rm p},3{\rm p}}\propto \frac{1}{t_{12}\ln\left(1+\eta_{2{\rm p},3{\rm p}}(t_{12}+t_{13})\right)\left(\frac{5}{2}+k_{res}(t_{12}+t_{23})\right)^{2}}.   
\label{eq:EchoesTotalIntegr}
\end{eqnarray} 
Here one can determine two distinguishable behaviors (cf. Fig. \ref{fig:Fig3TLS}) depending whether the characteristic times are shorter (intermediate regime) or longer (long time limit)  than the thermal or resonant TLS relaxation times $1/k_{T}, 1/k_{res}$. In the intermediate regime the echo signal is due to typical  symmetric resonant TLS, $q\ll 1$, $|\widehat{\gamma}|\approx \gamma_{0}$. The power law time dependent echo decay is defined by the phase volume of these TLS $\frac{qt_{12}(t_{12}+t_{23})k_{T}TP_{0}U_{0}}{\hbar}\leq 1$ which results in the laws  $t_{12}^{-2}$ or $(t_{12}t_{13})^{-1}$ for two and three pulse echoes, respectively. At longer times the echo decay is determined by the relaxation and decoherence free subset of TLS weakly coupled to phonons, Tr$(\widehat{\gamma}^2)|\gamma_{0}|^{-2}\leq \frac{1}{k_{{\rm res}}t_{23}}$. Their phase volume decreases with the time as $t_{23}^{-2}$ which determines the law of echo decay close to the experimentally observed one, Fig. \ref{fig:Fig3TLS}. One should notice that three pulse echo data cannot be fitted by the exponential  decay occurring if the distribution of the TLS-phonons coupling constant, Eq. (\ref{eq:gammadistr}), is ignored. The crossover between the two quasi-power law regimes is used below to estimate the TLS relaxation rates. 

One should notice that the ``longitudinal" part of TLS elastic tensor cannot be ignored in the very long time limit, $t_{23} \gg 1/k_{{\rm res}}$. It leads to an extra factor $1/\sqrt{5/2+k_{l}(t_{12}+t_{23})}$, where $k_{l} \sim k_{{\rm res}}(v_{t}/v_{l})^5 \sim 0.1k_{{\rm res}}$ for the phonon stimulated relaxation  and changes the long time asymptotic behavior from $I\propto t_{13}^{-2}$ to $t_{13}^{-2.5}$ at $t_{13}>10/k_{{\rm res}}$. Since in the experiment under consideration the times of measurement are shorter (see Fig. \ref{fig:Fig3TLS} and the relaxation time estimates) the longitudinal part of elastic tensor is not significant.

Theoretical model is expressed by  Eq. (\ref{eq:EchoesTotalFit}) and experimental data are available for two and three pulse echoes in Borosilicate glass BK7\cite{thesis1,Gudrun} for the wide range of temperatures including $T=7.5$, $15$, $25$, $40$, $55$, $70$ and $100$mK for two pulse echoes ($1\mu$s$<t_{12}\leq 1$ms) and $T=7.5$, $15$, $25$, $40$, $55$, $70$mK and durations $t_{12}=2$, $5.5$ and $15\mu$s, for three pulse echoes ($6\mu$s$<t_{23}\leq 10$ms). Applying simple Monte-Carlo algorithm minimizing the mean squared deviations between logarithms of experimental data and theoretical fits to find an optimum set of the free parameters defined above (see Appendix for detail), we obtained a good fit of all experimental data choosing $A=4.66\cdot 10^{7}$s$^{-1}$K$^{-3}$, $B=6.3\cdot 10^{4}$s$^{-1}$K$^{-1}$, $\alpha=0.018$, $\xi=0.46$, $\eta_{2{\rm p}}=0.63$, $\eta_{3{\rm p}}=9.45$. Some examples of comparison of theory vs. experiment for two-pulse and three pulse echo time dependencies are shown in Figs. \ref{fig:Fig2TLS} and \ref{fig:Fig3TLS}. For all data sets the accuracy of the fit is close to the accuracy of the experiment (see Appendix). 

Let us compare our fitting parameters with available experimental data for BK7. The latter three parameters $\xi=0.46$, $\eta_{2{\rm p}}=0.67$, $\eta_{3{\rm p}}=9.45$, are of order of unity as expected. Difference between parameters $\eta_{2p}=0.63$ and $\eta_{3p}=9.45$ can be because of the different scenarios of decoherence for two and three pulse echoes (see Appendix). The more accurate study of the problem requires the application of the telegraph process formalism \cite{telegraph} which is outside of the scope of the present paper. 

The first three parameters, $A$, $B$ and $\alpha$ can be directly compared with the results of the previous investigations of BK7. For convenience we assign the subscript ``echo" to our Monte-Carlo estimates, while the estimates based on previous studies are given with the subscript ``p". According to the classical paper\cite{Jackle72} one can express the parameter $A_{p}$ from Eq. (\ref{eq:DiagRelRate}), in terms of the  TLS coupling strengths to longitudinal, $\gamma_{l}$, and transverse, $\gamma_{t}$, phonons having velocities $c_{l}$ and $c_{t}$, respectively, as
\begin{equation}
A_{p}=\left(\frac{\gamma_{l}^{2}}{c_{l}^{5}}+\frac{2\gamma_{t}^{2}}{c_{t}^{5}}\right)\frac{k_{B}^3}{2\pi\hbar^{4}\rho}. 
\label{eq:PhEmissionRate}
\end{equation} 
Using the earlier experimental results for BK7 \cite{Doussineau1983} we get $A_{p}\approx 4.0\cdot 10^{7}$s$^{-1}$K$^{-3}$, which is consistent with the estimate based on the two and three pulse echo  experimental data, $A_{echo}=4.66\cdot 10^{7}$s$^{-1}$K$^{-3}$. Some overestimate of the relaxation rate can be due to the small effect of longitudinal part of TLS-phonon interaction and/or dipole-dipole TLS interaction, as noticed above.

Consider the relative strength of TLS-TLS interaction, $\chi=P_{0}U_{0}$. It is defined by the fitting parameter $\alpha_{{\rm echo}}=\frac{\pi^6}{24}\chi=0.018$ as $\chi_{{\rm echo}} \approx  0.44\cdot 10^{-3}$. Alternatively, one can estimate the average TLS-TLS interaction constant $U_{0p}$ using the TLS coupling with phonons, $\widehat{\gamma}_{p}$ determined by  the  relaxation parameter $A_{p}$. Applying the Fermi golden rule to the symmetric TLS relaxation we get
$A_{p}=\left(\frac{2}{15v_{l}^{5}}+\frac{1}{5v_{t}^{5}}\right)\frac{|\gamma_{\rm p}|^2 k_{B}^3}{8\pi\hbar^{4}\rho}$. 
Comparing this expression with the above estimates of the parameter $A_{{\rm echo,p}}$ one can find the average elastic tensor squared invariant as\cite{Doussineau1983} $|\gamma_{{\rm p}}|=4.18$eV.
 TLS coupling constant, $U_{0}$, is defined as a quadratic form of two TLS  elastics tensors depending on their orientations. Assuming $\widehat{\gamma}$'s to be traceless random Gaussian matrices with $|\widehat{\gamma}|$ obeying the statistics, Eq. (\ref{eq:gammadistr}), we evaluated $U_{0}$ numerically and obtained $\chi_{{\rm p}}\approx 0.5\cdot 10^{-3}$ consistently with the echo based estimate. 

Consider the TLS-TLS interaction contribution to the TLS relaxation rate characterized by the parameter $B$ (see Eq. (\ref{eq:DiagRelRate})). Using the previously estimated dimensionless parameter $\chi$ and the qualitative expression of Ref. \cite{abreview}, $B_{{\rm th}}=10 k_{B}\chi^3/\hbar$K$^{-1}$s$^{-1}$, we get $B_{{\rm th}}\approx 164$K$^{-1}$s$^{-1}$, which is smaller than  the echo based estimate, $B_{{\rm echo}}\sim 64000$, by the factor of $380$. To attain an agreement between experiment and theory  one should set $\chi_{th}\approx 0.0035$ which is a factor of $7$ larger than the estimate above, $\chi \sim 0.0005$. 

What is the source of this discrepancy? One possible explanation is that there is a different TLS relaxation  mechanism which leads to the faster relaxation than the one proposed in Refs.\cite{abrel,abreview} Another reason could be that the qualitative theory\cite{abreview} misses a large numerical parameter, which can account for the above discrepancy. This looks quite realistic if we consider how the parameter $\chi$ enters other theories. For instance in the expression for the spectral diffusion induced phase decoherence, Eq. (\ref{eq:EchoesTheorySingleTLS}), it enters together with the large numerical factor $\pi^{6}/24\sim 40$. This factor comes from the unity sphere volume contribution and  integration over energies. One can expect a similar rise of the effective parameter $\chi$ in the theory of interaction stimulated relaxation, which is more than sufficient to interpret the present experimental data. Of course, it is desirable to develop a more rigorous theory of the TLS-TLS interaction stimulated relaxation, which is beyond the scope of the present paper. 

Alternatively the theory can be tested by the experimental verification of the relaxation rate parametric dependence $k \propto \chi^3$. For instance, using Ref. \cite{Doussineau1983} one can estimate  $\chi \sim 0.00042$ for $a-$SiO$_{2}$. Accordingly the interaction stimulated relaxation rate in $a-$SiO$_{2}$ should be smaller than that in BK7 by a factor of $2$, which is interesting to verify experimentally. Preliminary analysis of the short time two-pulse echo measurements performed using the same experimental setup \cite{Gudrun} yields $B_{{\rm SiO}_{2}} \approx 22000 \approx B_{{\rm BK7}}/3$ which is approximately consistent with our expectation.

Thus we proposed a general theory interpreting two and three pulse dielectric echo measurements at microwave frequency, $\omega_{0}$, in low temperature amorphous solids. We show that  at long  times the echo amplitude is determined by the relaxation and decoherence free subset of symmetric TLS having small coupling with phonons, Tr$(\widehat{\gamma}^2)\sim 1/t_{23}$, with the phase volume approximately scaling as $1/t_{23}^2$. This power law describes the decay of the echo with the time in the long time limit. Our theory interprets the recent experimental data in BK7 for two and three pulse echoes within the experimental accuracy. One can conclude that the dielectric echo measurements can be used to determine the TLS relaxation rate based on the crossover between different power law time dependencies rather than the exponential decays.  

The relaxation rate temperature dependence satisfying the experimental data can be represented as a superposition of two contributions due to phonons, $k_{ph}\propto T^{3}$, and due to TLS-TLS interaction, $k_{TLS}\propto T$. This observation  is consistent with previous measurements. The relaxation rate due to TLS - phonon interaction extracted from the echo experimental data is consistent  with the previous measurements in BK7. The nature of the quantitative discrepancy between experiment and theory for TLS-TLS interaction stimulated relaxation in glasses is discussed and further experimental verifications of that theory are proposed. 


We acknowledge Moshe Schechter for useful discussions and hospitality during ab visit to Israel and Ora Entin-Wohlman and Amnon Aharony for useful discussion of the idea of decoherence free two level system developed in their earlier work.\cite{Ora} ab and JL acknowledge the hospitality and partial support of their visit by the Heidelberg University and by the LINK Program of NSF and Louisiana Board of Regents, Award. no. NSF(2012)-LINK-65.  This work is partially supported  by the NSF EPSCoR LA-SiGMA project, awards no. EPS-1003897 and the European Community Research Infrastructures under the FP7 Capacities Specific Programme, MICROKELVIN project number 228464.

\section{Appendix}
\label{sec:App}

\subsection{Echo decay  due to the spectral diffusion}
\label{sec:decoh}

Here we characterize the echo decay associated with the TLS phase fluctuations caused by its interaction with neighboring TLS.\cite{BlackHalperin} This decay is defined by the following average function (for two-pulse echo one should set $t_{23}=t_{12}$) 
\begin{eqnarray}
\left<\exp\left(\frac{i}{\hbar}\int_{0}^{t_{12}}d\tau (E(\tau)-E(\tau+t_{23}))\right)\right>. 
\label{eq:echoDecayCommon}
\end{eqnarray}
where $E(\tau)$ is the time-dependent energy of the TLS under consideration having average asymmetry $\Delta$ and elastic interaction tensor $\widehat{\gamma}$. If a TLS energy is time independent, then two phases in the exponent cancel each other and no decay occurs. The time dependent part of TLS energy can be represented as\cite{BlackHalperin,abreview} 
\begin{eqnarray}
\delta E(t) = qx\sum_{j} U_{j}S^{z}_{j}(t),
\nonumber\\
q=\frac{|\Delta|}{E}, ~ x = \frac{|\widehat{\gamma}|}{\gamma_{0}},   
\label{eq:EnergyFluct}
\end{eqnarray} 
where $qxU_{j}$ is the interaction of the given TLS with some TLS $j$, factors $q$ and $x$ are separated to have the bare interaction $U_{j}$ independent of the TLS under consideration similarly to Refs. \cite{BlackHalperin}. Pseudospin $1/2$ operators $S^{z}(t)$ describes the present states of neighboring TLS being $1/2$ in the ground state and $-1/2$ in the excited state. The spontaneous transitions between these states determine the time dependence of TLS energy.

Then ignoring correlations between weakly interacting TLS one can represent the phase factor, Eq. (\ref{eq:echoDecayCommon}), as
\begin{eqnarray}
\left<\exp\left(\frac{i}{\hbar}\int_{0}^{t_{12}}d\tau (E(\tau)-E(\tau+t_{13}))\right)\right>
\nonumber\\
=\left<\prod_{j}\exp\left(\frac{ixqU_{j}}{\hbar}\int_{0}^{t_{12}}d\tau (S^{z}_{j}(\tau)-S^{z}_{j}(\tau+t_{23}))\right)\right>. 
\label{eq:echoDecayCommon1}
\end{eqnarray}
Using the general method of averaging the products over neighbors\cite{BlackHalperin,abreview} characterized by the density of states $P(E_{1}, \Delta_{01})$ 
\begin{eqnarray}
<\prod_{i}F(i)>=\prod_{\mathbf{R}}(1-d\mathbf{R}\int_{0}^{\infty}dE_{1}\int_{0}^{E_{1}}d\Delta_{01}P(E_{1}, \Delta_{01})(1-F(E_{1}, \Delta_{01}, \mathbf{R}))=
\nonumber\\
=\exp\left[-\int d\mathbf{R}\int_{0}^{\infty}dE_{1}\int_{0}^{E_{1}}d\Delta_{01}P(E_{1}, \Delta_{01})(1-F(E_{1}, \Delta_{01}, \mathbf{R}))\right] 
\label{eq:GenAvProd}
\end{eqnarray}
one can reexpress Eq. (\ref{eq:echoDecayCommon1}) as  $\exp\left[-qxG(t_{12}, t_{23})\right]$ where the function $G$ is defined as 
\begin{eqnarray}
 G(t_{12}, t_{23})=\chi\int d\mathbf{x} \int_{0}^{\infty}\frac{d\Delta_{01}}{\Delta_{01}} \int_{\Delta_{01}}^{+\infty} dE_{1}\left(1-\left<\exp\left[\frac{i}{2\hbar x^3}\int_{0}^{t_{12}}d\tau (S^{z}_{1}(\tau)-S^{z}_{1}(\tau+t_{12}))\right]\right>\right),
\nonumber\\
\chi = P_{0}U_{0}.  
\label{eq:echoesTheoryExp1}
\end{eqnarray}
Here the replacement of variable $x^{3}=\frac{|\Delta\Delta_{1}|}{E_{1}|u_{1}|}R^3$ has been used. This substitution results in the prefactor of the average absolute value of TLS coupling constant, $U_{0}$.

In the case $t_{12}\ll t_{23}, k_{s}^{-1}$ realized in the majority of three pulse echo measurements  one can simplify the integral in the exponent in Eq. (\ref{eq:echoesTheoryExp1}) as $\int_{0}^{t_{12}}d\tau (\delta S^{z}(\tau)-\delta S^{z}(\tau+t_{23}))\approx t_{12}(S^{z}(0)-S^{z}(t_{23}))$.  The exponent can be then evaluated explicitly using the properties of the spin $1/2$ operator, $e^{iaS^{z}}=\cos(a/2)+2iS^{z}\sin(a/2)$, and the spin-spin correlation function $<\delta S^{z}(t)\delta S^{z}(0)>=\frac{e^{-kt}}{4\cosh^{2}\left(\frac{E}{2k_{B}T}\right)}$, where $k=\left(\frac{\Delta_{0}}{E}\right)^{2}k_{s}(E, T)$ is a TLS relaxation rate and $k_{s}(E, T)$ is a  relaxation rate at temperature $T$ for symmetric TLS ($\Delta=0$) with energy $E$. Using these definitions one can express Eq. (\ref{eq:echoesTheoryExp1}) in the form 
\begin{eqnarray}
G(t_{12}, t_{23})=\chi\int d\mathbf{x} \left(1-\cos\left(\frac{t_{12}}{x^{3}}\right)\right)\int_{0}^{1}\frac{dv}{v} \int_{0}^{+\infty} \frac{dE_{1}}{4\cosh^{2}\left(\frac{E_{1}}{2k_{B}T}\right)}\left(1-e^{-v^{2}k_{s}(E_{1}, T)t_{23}}\right),
\nonumber\\
v=\frac{\Delta_{01}}{E_{1}}.  
\label{eq:echoesTheoryExp2}
\end{eqnarray} 
The integral over $x$ can be evaluated exactly as $\int d\mathbf{x}\left(1-\cos\left(\frac{t_{12}}{x^{3}}\right)\right)=\frac{2\pi^2}{3}|t_{12}|$. Then, after the replacement of variable, $u=\frac{E_{1}}{2k_{B}T}$, we get 
\begin{eqnarray}
G(t_{12}, t_{23})=\frac{2\pi^2}{3\hbar}\chi t_{12}k_{B}T\int_{0}^{1}\frac{dv}{v} \int_{0}^{+\infty} \frac{du}{\cosh^{2}(u)}\left(1-e^{-v^{2}k_{s}(2uk_{B}T, T)t_{23}}\right).  
\label{eq:echoesTheoryExp3}
\end{eqnarray} 
The analytical evaluation of the integrals in Eq. (\ref{eq:echoesTheoryExp2}) is not possible. There are two asymptotic behaviors for long and short delay times $t_{23}$ compared to the thermal TLS relaxation time $T_{1T} \sim k_{s}(k_{B}T, T)^{-1}$, namely. 
\begin{eqnarray}
G(t_{12}, t_{23})\approx\left\{\begin{array}{cl}
	\frac{t_{12}U_{T}}{\hbar}k_{T1}t_{23}, & k_{s}(k_{B}T, T)t_{23} < 1, \\
	\frac{t_{12}U_{T}}{\hbar}\ln (k_{s}(k_{B}T, T)t_{23}), & k_{s}(k_{B}T, T)t_{23} > 1,
	   \end{array}\right.
\nonumber\\
U_{T}= \frac{\pi^2}{3}k_{B}TP_{0}U_{0}, ~ k_{T1}=\int_{0}^{+\infty} \frac{k_{s}(2uk_{B}T, T)du}{\cosh^{2}(u)}. 	   
\label{eq:echoesTheoryExpAsympt}
\end{eqnarray} 
According to the earlier studies  \cite{Jackle76,abrel,abreview,EnssReview,Esquinazi,OsheroffRelaxation,Classen,Ladieu,Fefferman}  TLS relaxation rate in dielectric glasses can be approximately represented as the sum of two contributions 
\begin{eqnarray}
k_{s}(E, T)=A\left(\frac{E}{k_{B}}\right)^{3}\coth\left(\frac{E}{2k_{B}T}\right) + BT, 
\label{eq:DiagRelRateSym}
\end{eqnarray}
associated with the TLS-phonon and TLS-TLS interactions, respectively. The phonon contribution to $k_{T1}$ can be evaluated analytically\cite{BlackHalperin} as $k_{T1ph}=\frac{\pi^4}{8}AT^3$, while we represent the average TLS-TLS interaction contribution in the form $k_{T1int}=\frac{\pi^4}{8}\xi AT^3$ with some unknown fitting parameter of the theory, $\xi\sim 1$, depending on the specific of the interaction stimulated relaxation.  

\begin{figure}[h!]
\centering
\includegraphics[width=8cm]{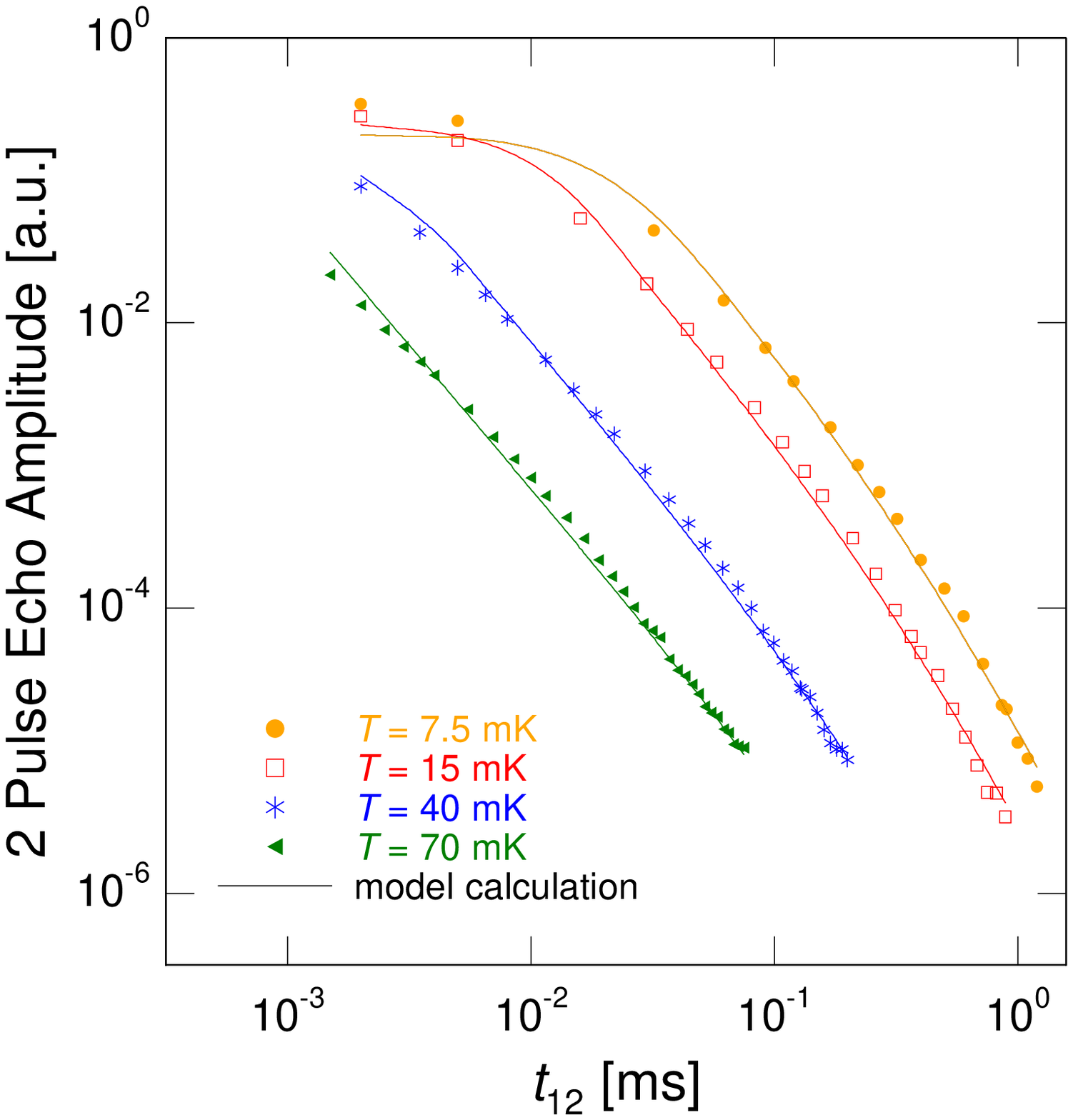}
\caption{ Experiment (symbols) vs. theory (lines) for the time dependence of the two pulse echo amplitudes in BK7. Straight lines (colored online) from the top to the bottom corresponds to experimental temperatures in ascending order, $7.5$mK, $15$mK, $40$mK and $70$mK, respectively.}
\label{fig:Fig2pTLSApp}
\end{figure}

\begin{figure}[ht]
\centering
\begin{center}$
\begin{array}{cc}
\includegraphics[width=6cm]{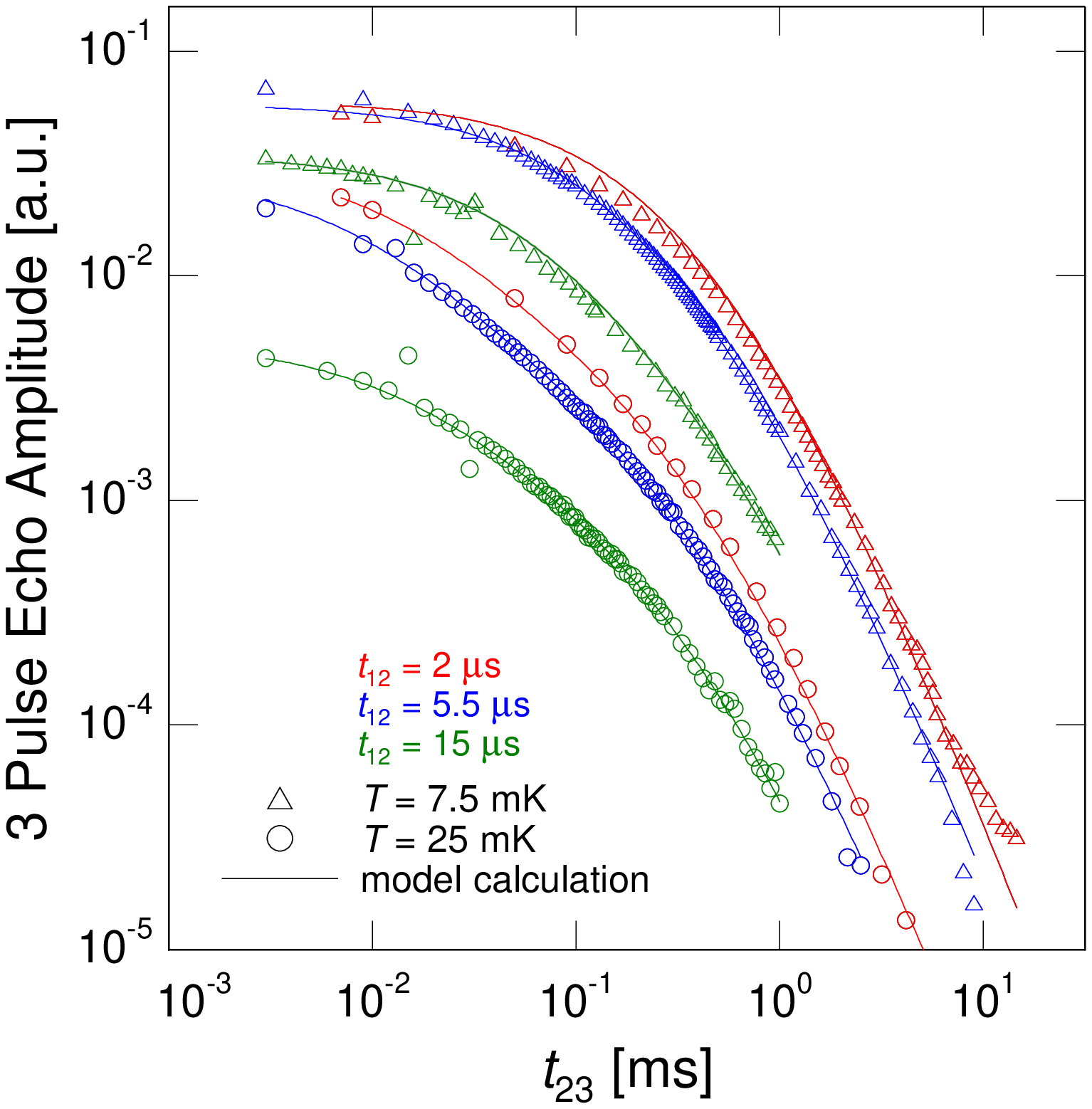} & 
\includegraphics[width=7cm]{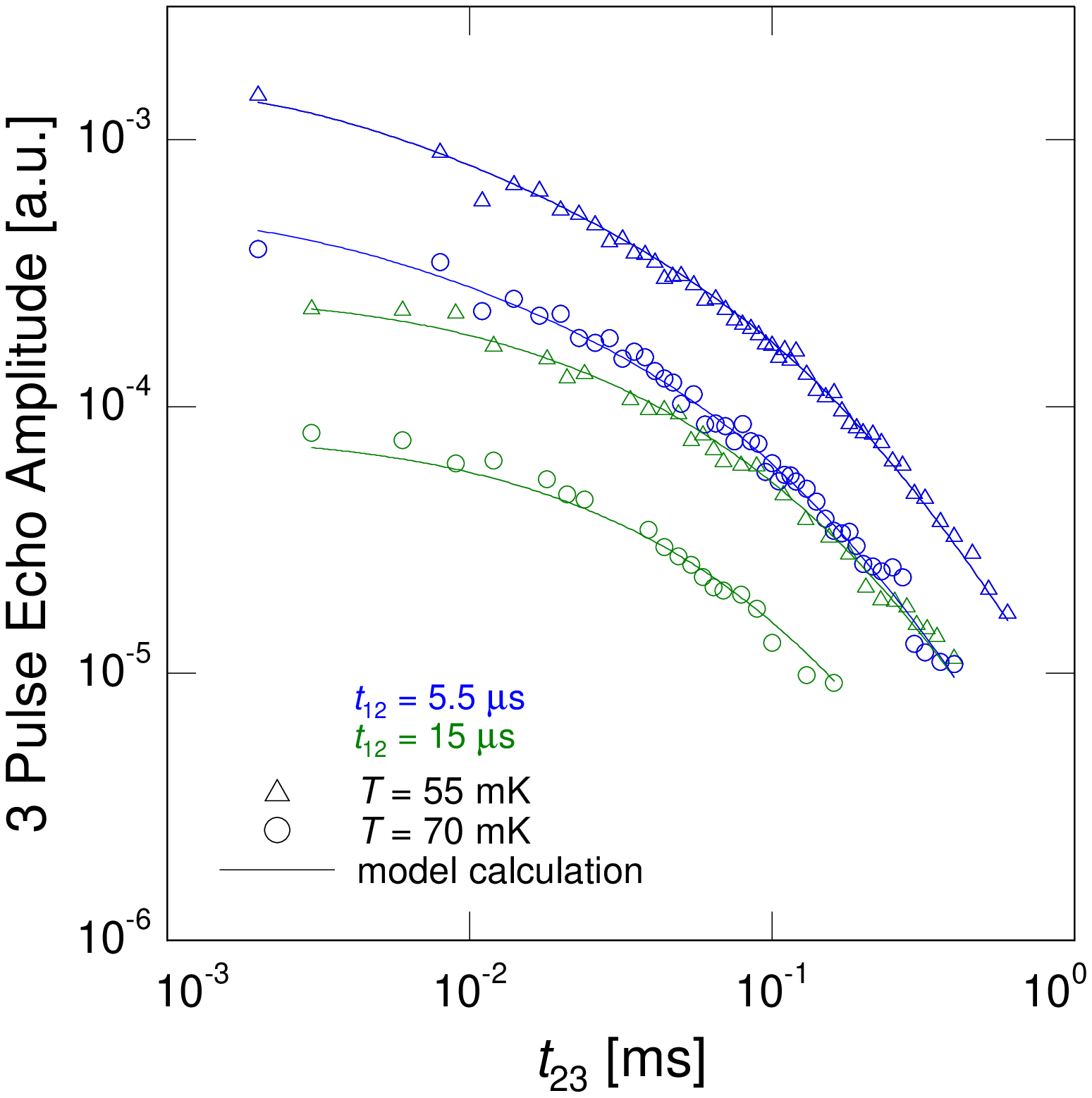} 
\end{array}$
\end{center}
\caption{Experiment (symbols) vs. theory (lines) for the time dependence of the three pulse echo amplitudes in BK7 ({\bf left}) for lower temperatures $T=7.5$mK and $T=25$mK and ({\bf right}) for higher temperatures $T=55$mK and $T=70$mK. Straight lines (colored online) from the top to the bottom corresponds to experimental temperatures in ascending order and different times $t_{12}$.)}
\label{fig:Fig3pTLSApp}
\end{figure}

For the analysis of experimental data we used simplified  analytical expressions 
\begin{eqnarray}
G_{2{\rm p}}(t_{12})=\frac{\alpha k_{B}T t_{12}}{\hbar}\frac{\ln\left(1+\eta_{2{\rm p}}k_{T}t_{12}\right)}{\eta_{2{\rm p}}};
\nonumber\\
G_{3{\rm p}}(t_{12}, t_{23})=\frac{\alpha k_{B}Tt_{12}}{\hbar}\frac{\ln\left(1+\eta_{3{\rm p}}k_{T}(t_{12}+t_{23})\right)}{\eta_{3{\rm p}}};
\nonumber\\
k_{T}=AT^{3}+ \xi BT. ~ \alpha=\frac{\pi^{6}}{24} P_{0}U_{0}.
\label{eq:EchoesTheorySingleTLSApp}
\end{eqnarray}
which handles all asymptotic behaviors for the three pulse echo, Eq. (\ref{eq:echoesTheoryExpAsympt}), and should be valid at least qualitatively for the two pulse echo (cf. Ref. \cite{BlackHalperin}).  The time $t_{23}$ in Eq. (\ref{eq:echoesTheoryExpAsympt}) is replaced with the sum $t_{12}+t_{23}$, which can be justified by the exact perturbation theory\cite{BlackHalperin} at intermediate times $t_{23}<k_{T}^{-1}$. To emphasize the difference between two cases we introduced two different free parameters $\eta_{2{\rm p}}$ and $\eta_{3{\rm p}}$ in two and three pulse echo expressions for the decay rates in Eq. (\ref{eq:EchoesTheorySingleTLS}). Unknown parameters $\alpha$, $\eta_{2, 3{\rm p}}$, $A$, $B$, $\xi$ are all used as free parameters of the theory determined by the best fit of the experimental data. Here we show all experimental data skipped in the main body of the paper together with theoretical curves for two pulse  (Fig. \ref{fig:Fig2pTLSApp}) and three pulse (Fig. \ref{fig:Fig3pTLSApp}) echo studies at different temperatures to illustrate nearly perfect agreement between the experiment and the theory.

\subsection{Monte-Carlo search for the free parameters}

As an input we have the sequence of experimental data $y_{i}$ for two and three pulse echoes in a wide range of temperatures and times of measurements to be fitted by the theoretical expression,  $z_{i}(\mathbf{P})$ (Eq. (8) in the main text), depending on the set of unknown fitting parameters $\mathbf{P}$. In each step of the procedure we calculate the logarithmic root mean squared deviation of experiment and theory 
\begin{equation}
D(\mathbf{P})=\sqrt{\sum_{i}(y_{i}-c\cdot z_{i}(\mathbf{P}))^2y_{i}^{-2}}.
\label{eq:logdev}
\end{equation}
The scaling constant $c$ is determined by the minimum requirement for the function Eq. (\ref{eq:logdev}). The logarithmic deviation criterion has been chosen because experimental echo amplitudes change several orders of magnitude depending on the time of measurements. 

Then the new set of parameters, $\mathbf{P}'$, is created multiplying each parameter by the random factor $R=e^{\psi x}$ with a random number $x$ distributed uniformly between $-1$ and $1$. Initially we choose $\psi=2$. For this new set of parameters  the new deviation $D'$  is calculated using Eq. (\ref{eq:logdev}). In case of $D'<D$ we took $\mathbf{P}'=\mathbf{P}$, or left $\mathbf{P}$ without changes otherwise. The procedure is repeated $1000$ times. Then we reduce $\psi$ by the factor of $2$ and repeated the search again. This procedure is continued until $\psi=0.125$ and the result is considered as a final answer for the set of optimum parameters. We applied this algorithm several times obtaining the final answers different by no more than $10\%$ from each other. 

The application of this method yields an accurate fit of all experimental data. Most strong deviations occurs for the two-pulse echo at short times $t_{12} \sim 1\mu$s. Perhaps this is because the driving field amplitude is not so small and perturbation theory used to derive Eq. (8) in the main text is not quite applicable. This is not important for intermediate and long times where only ``symmetric" TLS contribute to echo, where the pulses related part of echo amplitude is decoupled from its decay. 

\end{document}